%% file: main.tex
\documentclass[%
 prc,
 aps,%
 amsmath,amssymb,
 reprint,%
]{revtex4-2}

\usepackage{graphicx}
\usepackage{dcolumn}
\usepackage{bm}
\usepackage[dvipsnames]{xcolor}
\usepackage{xspace}
\usepackage{hyperref}
\usepackage{tikz}
\usetikzlibrary{calc}
\usepackage{siunitx}
\usepackage [latin1]{inputenc}

\usepackage{soul}

\usepackage[toc,page]{appendix}

\usepackage{algorithm}
\usepackage{algorithmicx}
\usepackage{algpseudocode}

\usepackage{siunitx}
\usepackage{enumitem}
\usepackage{balance}

\begin{document}

\preprint{AIP/123-QED}

\input{macro}
\title[Sample title]{Effectiveness of denoising diffusion probabilistic models for fast and high-fidelity\\ whole-event simulation in high-energy heavy-ion experiments}

\author{Dmitrii Torbunov}
\thanks{Contributed equally to this work}
\affiliation{Computational Science Initiative, Brookhaven National Laboratory, Upton, New York 11973}

\author{Yeonju Go}
\thanks{Contributed equally to this work}
\affiliation{Physics Department, Brookhaven National Laboratory, Upton, New York 11973}

\author{Timothy Rinn}
\affiliation{Physics Department, Brookhaven National Laboratory, Upton, New York 11973}

\author{Yi Huang}
\affiliation{Computational Science Initiative, Brookhaven National Laboratory, Upton, New York 11973}

\author{Haiwang Yu}
\affiliation{Physics Department, Brookhaven National Laboratory, Upton, New York 11973}

\author{Brett Viren}
\affiliation{Physics Department, Brookhaven National Laboratory, Upton, New York 11973}

\author{Meifeng Lin}
\affiliation{Computational Science Initiative, Brookhaven National Laboratory, Upton, New York 11973}

\author{Yihui Ren}
\affiliation{Computational Science Initiative, Brookhaven National Laboratory, Upton, New York 11973}

\author{Jin Huang}
\affiliation{Physics Department, Brookhaven National Laboratory, Upton, New York 11973}
\email{jhuang@bnl.gov}


\begin{abstract}
Artificial intelligence (AI) generative models, such as generative adversarial networks (GANs), variational auto-encoders, and normalizing flows, have been widely used and studied as efficient alternatives for traditional scientific simulations. However, they have several drawbacks, including training instability and inability to cover the entire data distribution, especially for regions where data are rare. This is particularly challenging for whole-event, full-detector simulations in high-energy heavy-ion experiments, such as sPHENIX at the Relativistic Heavy Ion Collider and Large Hadron Collider  experiments, where thousands of particles are produced per event and interact with the detector. This work investigates the effectiveness of Denoising Diffusion Probabilistic Models (DDPMs) as an AI-based generative surrogate model for the sPHENIX experiment that includes the heavy-ion event generation and response of the entire calorimeter stack. DDPM performance in sPHENIX simulation data is compared with a popular rival, GANs. Results show that both DDPMs and GANs can reproduce the data distribution where the examples are abundant (low-to-medium calorimeter energies). Nonetheless, DDPMs significantly outperform GANs, especially in high-energy regions where data are rare. Additionally, DDPMs exhibit superior stability compared to GANs. The results are consistent between both central and peripheral centrality heavy-ion collision events. Moreover, DDPMs offer a substantial speedup of approximately a factor of 100 compared to the traditional Geant4 simulation method.
Source code is available at \url{https://github.com/LS4GAN/calo-ddpm}.
\end{abstract}

\maketitle

\section{Introduction}

Simulating the interaction of particles with detectors in nuclear experiments is a computationally intensive task. The Monte Carlo (MC) method employed for this purpose involves tracking each particle's interaction with the detector material.
Software packages such as Geant4~\cite{GEANT4:2002zbu,Allison:2006ve,Allison:2016lfl} are commonly used, followed by digitization simulation. Calorimeter simulations are particularly time-consuming due to the large number of interaction steps induced by many secondary particles as they travel through the detector geometry. Simulating full-detector events with physics backgrounds can be even more computationally expensive. However, deep generative machine learning (ML) models have emerged as a promising alternative for faster simulations~\cite{Belayneh:2019vyx,Shanahan:2022ifi}.

Deep generative neural networks learn a data distribution and create new data
points implicitly sampling the learned distribution. There have been many
different techniques developed over the past two decades. Auto-encoders~\cite{hinton2006reducing} and variational auto-encoders~\cite{kingma2019introduction} 
task neural networks to encode or compress data into lower-dimensional features
in an embedding space and decode or expand the features back to the original
ones. They sample new data by first sampling in the learned embedding space and
decoding to the data space. Normalization flows~\cite{rezende2015variational,papamakarios2021normalizing} draw random samples from a simple distribution and transform
them to the target data distribution via a sequence of invertible and
differentiable mappings. Generative adversarial networks (GANs)~\cite{goodfellow2014generative} set up a minimax game for two neural networks, a generator and
a discriminator, competing against each other. The generator produces synthetic
data from random samples, while the discriminator discerns the synthetic data from
real ones. Once a 
balance is reached between the generator and the
discriminator, one can sample new synthetic data using the trained generator.
Recently, the Denoising Diffusion Probabilistic Model (DDPM)~\cite{ho2020denoising,SohlDickstein:2015dhe,arxiv190705600} and its variations~\cite{song2020denoising,song2020score} have drawn attention for their stability in
model training and high quality and diversity of the synthetic data. Meanwhile, contemporary text-to-image generative models~\cite{ramesh2022hierarchical}, integrated with large language models, have introduced impactful artificial intelligence (AI) to various scientific domains~\cite{kazerouni2023diffusion,corso2023diffdock}.

GANs have been successfully used to produce fast calorimeter simulations~\cite{Jaruskova:2023cke, Buhmann:2021vlp, Rehm:2021qwm, Khattak:2021xjl, Khattak:2021ndw, Erdmann:2018jxd, deOliveira:2017pjk, Paganini:2017dwg, Paganini:2017hrr}. Diffusion models in calorimeter and other detector simulations also have been gaining popularity~\cite{Amram:2023onf,Mikuni:2023tqg,Shmakov:2023kjj,Leigh:2023zle,Mikuni:2023dvk,Leigh:2023toe,Kansal:2020svm}. However, while most fast simulation algorithms are designed to handle single-shower simulations, research involving full-detector, whole-event ML simulations remains sparse. 
Full-detector simulations are essential in contexts such as heavy-ion collisions, where each event generates thousands of particles. They also play a key role in simulating the full-detector background, such as synchrotron radiation in detectors at the Electron-Ion Collider (EIC)~\cite{AbdulKhalek:2021gbh} where millions of photons will have to be simulated per event despite rare detector hits. Typically, a large background sample is required for detailed signal simulation embedding. Due to the high complexity and computational requirements of these simulations, they present a significant challenge to the scientific community. 

In high-energy heavy-ion collisions, there are highly nontrivial global correlations in the deposited energy, arising from phenomena like flow, minijets, or resonances/decays. 
Therefore, comprehensive, concurrent simulation of the entire event is imperative. Furthermore, numerous experiments measure rare and high-energy signals in the detector. This puts significant emphasis on the high-fidelity generation of rare features in both short- and long-distance scales, which poses a challenge for generative models. 

This work introduces the first full-detector, whole-event heavy-ion simulation using DDPM and GAN models trained on the simulated calorimeter data from the sPHENIX~\cite{PHENIX:2015siv, Belmont:2023fau} experiment.
The experimental background is discussed in Section II. The algorithm is explained in Section III, followed by the results and discussions in Sections IV and V, respectively.

\section{Problem Setting and Data Generation}

The reference heavy-ion dataset used in this work is produced using the HIJING~\cite{Wang:1991hta} MC event generator for Au+Au collisions at nucleon-nucleon center-of-mass energy (\snn) of $\SI{200}{GeV}$. The generated, long-lived particles then are subjected to a Geant4 simulation of the sPHENIX detector to model the effects of the detector. Events are generated with varying centrality and thereby exhibit distinct event characteristics. Specifically, two centrality ranges are used in this study: $0$--$10\%$ (referred to as ``central'' collisions) and $40$--$50\%$ (referred to as ``mid-central'' collisions), corresponding to larger and smaller medium sizes, respectively. In HIJING, the centrality is determined using the percentile of the impact parameter ($b$) between the colliding nuclei. The $b$-range of $0$--$4.88\ \mathrm{fm}$ ($9.71$--$10.81\ \mathrm{fm}$) corresponds to the centrality range of $0$--$10\%$ ($40$--$50\%$). In the generation, multiple simultaneous collisions in beam bunches, referred to as ``pileup events'', are allowed.  

\begin{figure}[t!]
  \begin{center}
    \resizebox{.98\linewidth}{!}{
\begin{tikzpicture}
    \def\lnpad{.3em}
    \node[inner sep=0] (left) at (0, 0) {
        \begin{tikzpicture}
            \node[anchor=north west, inner sep=0] (tower) at (0,0) {\includegraphics[height=360pt]{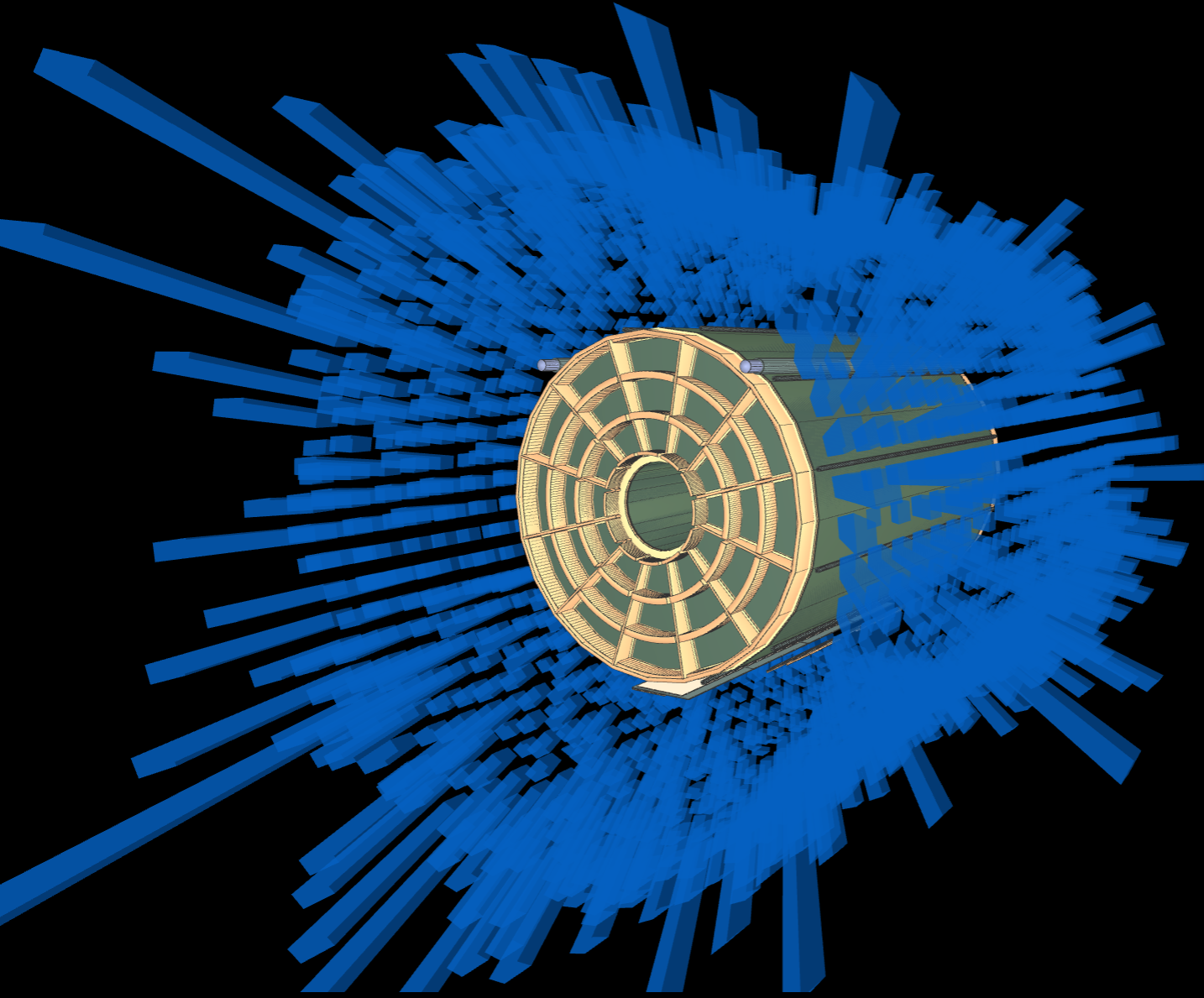}};
            \node[anchor=north west, inner sep=0, align=left, white, font=\fontsize{20}{20}\selectfont, inner sep=10pt] (text) at (tower.north west) {\bf{Hijing+Geant4 simulation}\\[\lnpad]\bf{sPHENIX geometry}\\[\lnpad]\bf{{Au}$+${Au} $\snn=\SI{200}{GeV}$}};    
        \end{tikzpicture}
    };
    \node[inner sep=0, anchor=north west] (right) at ($(left.north west)!1.03!(left.north east)$) {
        \begin{tikzpicture}
            \node[anchor=north west, inner sep=0] (tower) at (0,0) {\includegraphics[height=360pt]{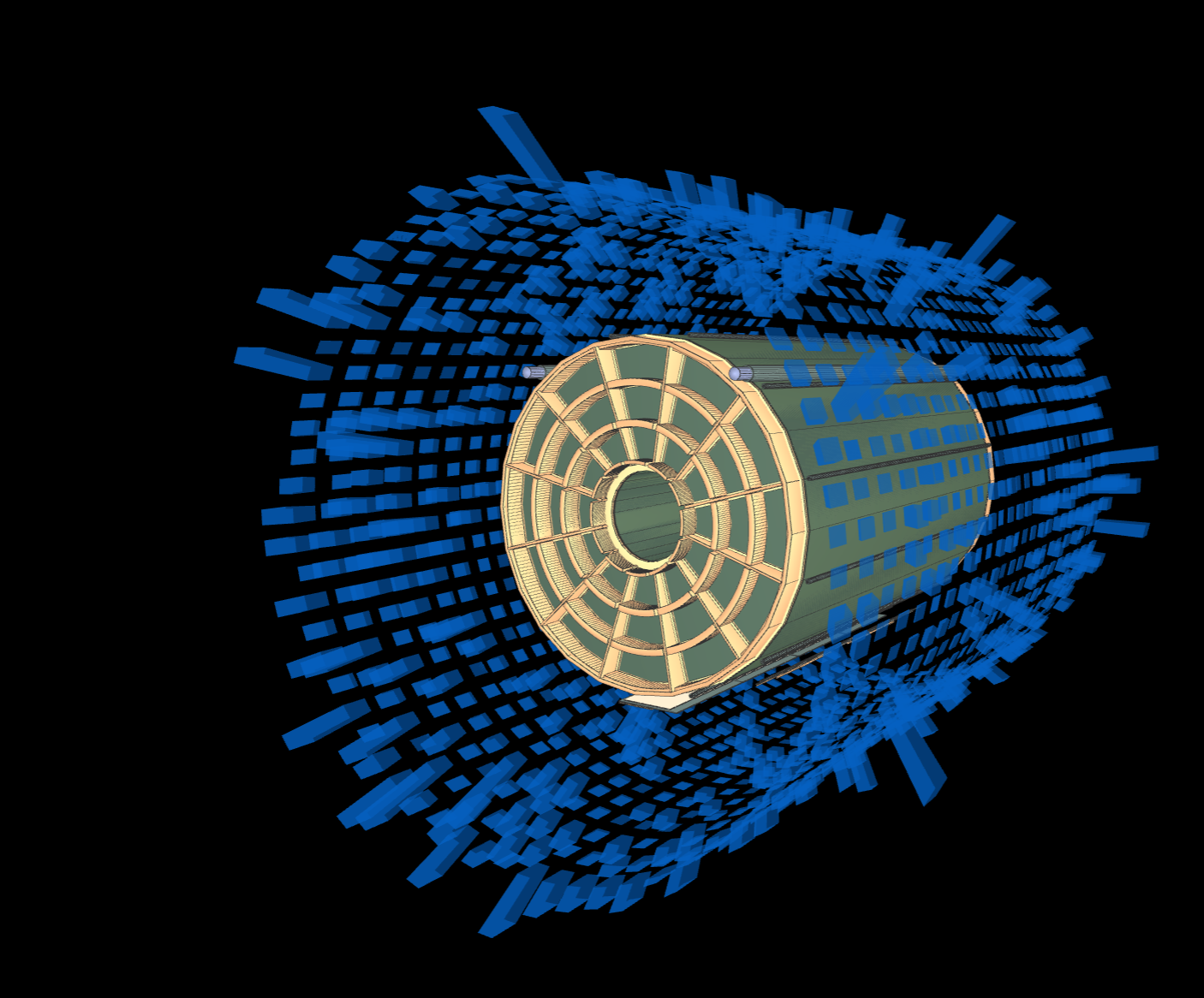}};
            \node[anchor=north west, inner sep=0, align=left, white, font=\fontsize{20}{20}\selectfont, inner sep=10pt] (text) at (tower.north west) {\bf{Hijing+Geant4 simulation}\\[\lnpad]\bf{sPHENIX geometry}\\[\lnpad]\bf{{Au}$+${Au} $\snn=\SI{200}{GeV}$}};    
        \end{tikzpicture}
    };
\end{tikzpicture}
}
    \caption{
    Display of tower energies (blue) deposited in the calorimeter system with the sPHENIX geometry for the $0$--$10\%$ (left) and $40$--$50\%$ (right) centrality ranges of example events. The center portion is an illustration of the tracking system that resides inside the calorimeters. 
    }
    \label{fig:sPHENIX_display}
  \end{center}
\end{figure}
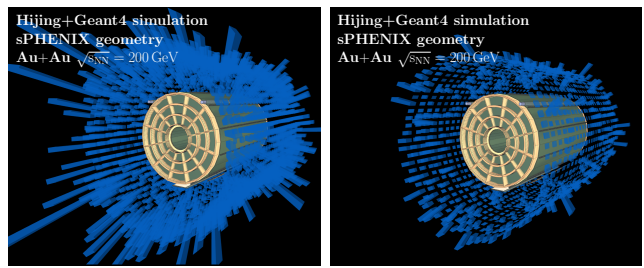

\begin{figure}[ht!]
\centering
    \def\width{.4\textwidth}
    \includegraphics[width=\width,trim=10 20 50 25,clip]{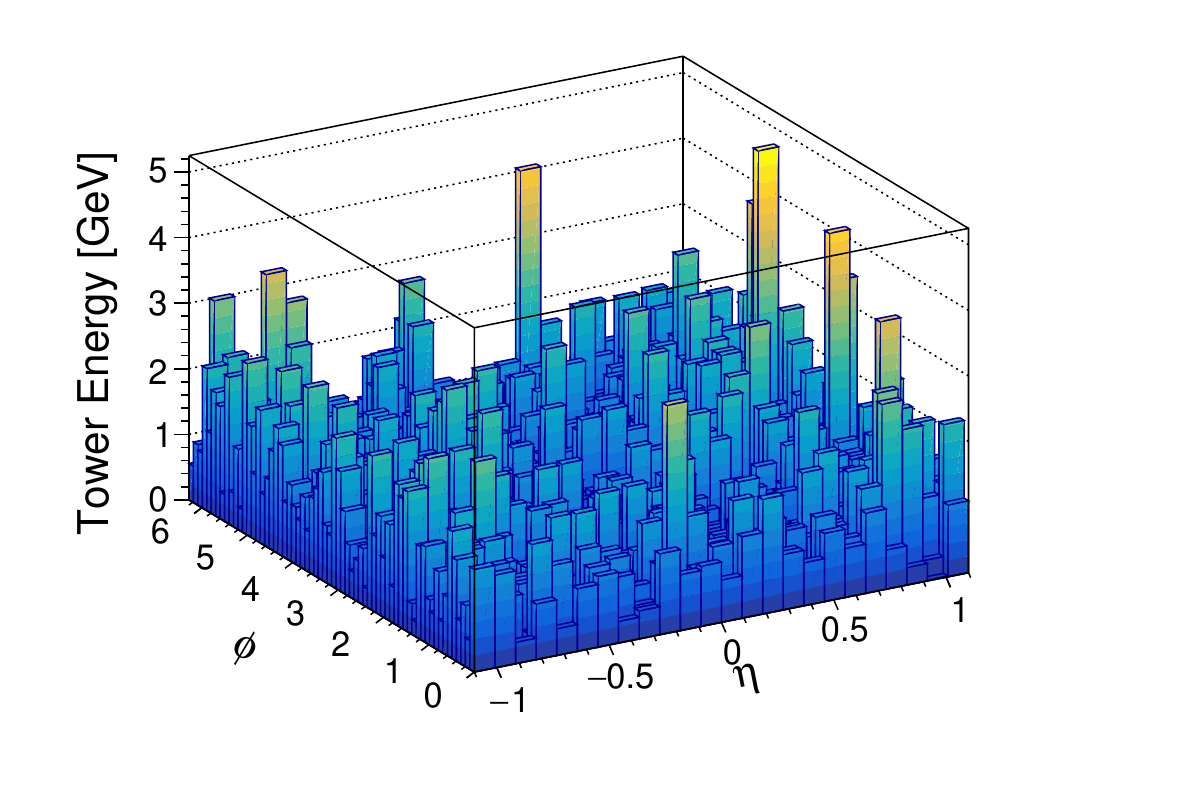}
    \includegraphics[width=\width,trim=10 20 50 25,clip]{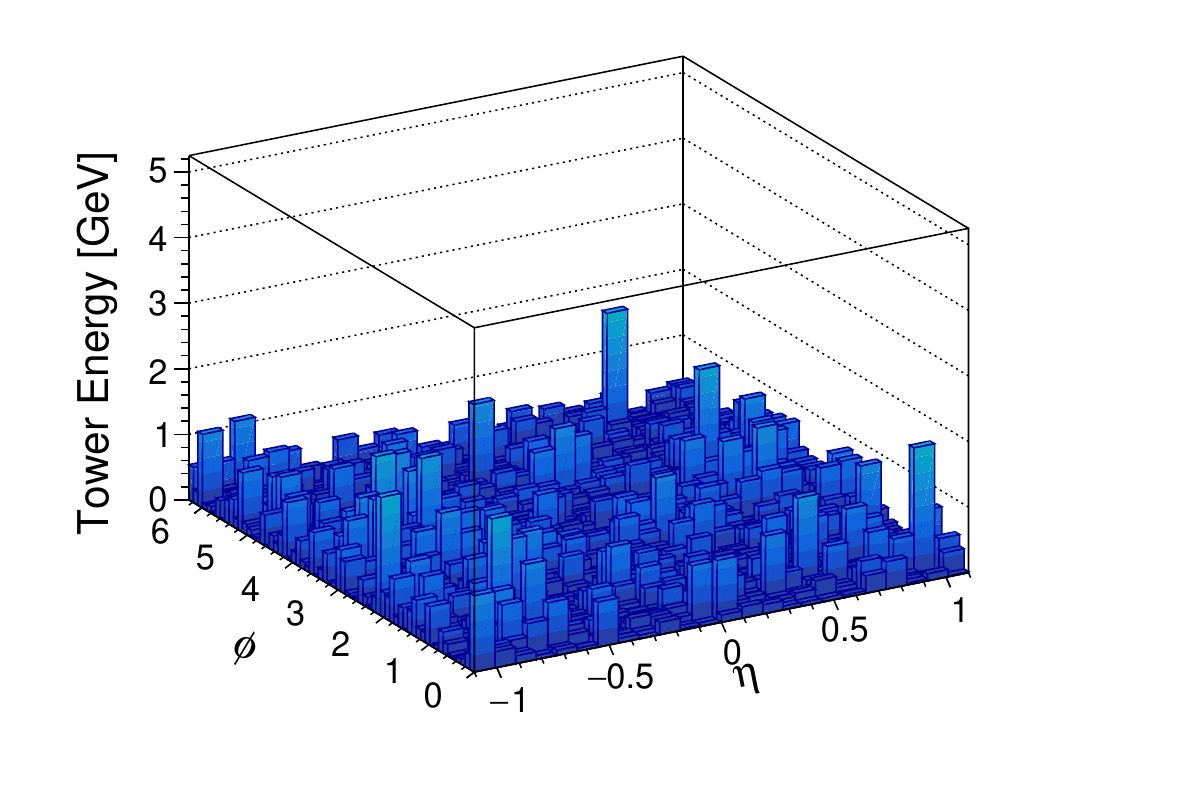}
    \caption{
    Tower energy distributions within the $(\eta, \phi)$-plane of the sPHENIX calorimeter system for the $0$--$10\%$ (top) and $40$--$50\%$ (bottom) centrality ranges of example events.
    }
    \label{fig:energy_eta_phi_map}
\end{figure}

The sPHENIX detector is equipped with three distinct calorimeters: electromagnetic (EMCal), inner hadronic (IHCal), and outer hadronic (OHCal)~\cite{sPHENIX:2017lqb}. These calorimeters possess a uniform and hermetic acceptance within the pseudo-rapidity ($\eta$) range of $-1.1 < \eta < 1.1$, covering the entire azimuthal angle ($0 < \phi < 2\pi$). Each calorimeter is segmented into distinct, sensitive volumes called \textit{towers}. The tower size of EMCal is $\deta\times\dphi = 0.025\times0.025$, while the tower sizes of IHCal and OHCal are $\deta\times\dphi = 0.1\times0.1$. Consequently, the EMCal tower size is four times finer than that of the IHCal/OHCal towers in both $\eta$ and $\phi$ dimensions. 
For a given event, the energies deposited within $4\times4$ EMCal towers (16 total towers) are aggregated with the energies deposited in individual IHCal and OHCal towers to match the tower sizes between the EMCal and HCal. This results in the formation of an integrated energy tower that comprises the three calorimeters. 
The number of integrated towers covered in these $\eta$ and $\phi$ ranges are 24 and 64, respectively. 
\autoref{fig:sPHENIX_display} shows the event displays of the combined tower energies with the sPHENIX geometry, generated by HIJING and simulated through Geant4 (\hijinggeant). Subsequently, this combined energy information is used to construct an energy image within the $(\eta,\phi)$-plane as shown in~\autoref{fig:energy_eta_phi_map} for the events of $0\,$--$10\%$ and $40\,$--$50\%$, respectively. The images containing (24, 64) towers in $(\eta,\phi)$ per event serve as the ground truth samples for training the ML models. Each centrality class is trained separately because event characteristics vary across centrality classes. Approximately 600,000 events are used for training in each centrality class.


\section{Diffusion Models}

\begin{figure}[t!]
\centering
    \def\width{.49\textwidth}
    \includegraphics[width=\width]{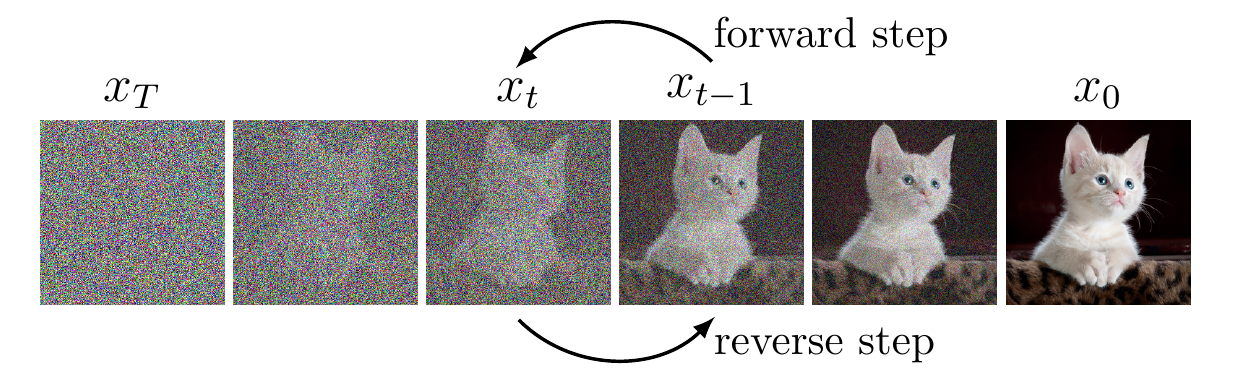}
    \caption{
    Illustration of the forward and reverse diffusion processes.
    The forward process ($\vx_0 \to \vx_T$) progressively adds Gaussian noise to an input image.
    The reverse process ($\vx_T \to \vx_0$) progressively denoises an image.
    }
    \label{fig:ddpm}
\end{figure}

Diffusion models are a class of generative models~\cite{yang2022diffusion}, capable of producing novel data from random noise.
This section reviews the DDPM model design and describes its training procedure.

\subsection{The DDPM Model}

The DDPM model~\cite{ho2020denoising} is built on the concept of a diffusion process (\autoref{fig:ddpm}) that has two directions: forward and reverse.
The forward diffusion process takes a sample from the data distribution $\vx_0 \sim p_\text{data}$ and transforms it to look like a sample from the standard normal distribution $\stdnorm$.
The forward process is defined through a set of $T$ iterations, so-called \textit{diffusion steps}, each of which adds a small amount of Gaussian noise to $\vx_0$.

The reverse process takes a sample from the standard normal distribution $\vx_T \sim \stdnorm$ and transforms it back to
the data distribution $p_\text{data}$.
Similar to the forward process, the reverse diffusion process is defined through a set of $T$ iterations, or \textit{sampling steps}, each of which
is an inverse of the corresponding forward iteration.
In effect, the reverse process removes the noise added in the forward steps. Thus, it is also referred to as the ``denoising process.''
\autoref{fig:ddpm} illustrates both processes.

The reverse process can be used for data generation. To generate a data sample, one starts with a random noise $\vx_T$ drawn from $\stdnorm$.
Then, one applies $T$ denoising iterations to $\vx_T$ and obtains a sample $\vx_0$.
If the reverse process is correct, then $\vx_0$ will be distributed according to $p_\text{data}$.
In other words, the reverse process can be considered as a generative map that transforms the standard normal distribution $\stdnorm$ into $p_\text{data}$.

Mathematically, the DDPM forward diffusion process is a Markov chain, where a single step is defined via~\autoref{eq:ddpm_fwd_step}~\cite{ho2020denoising}.
The forward process is parameterized by a set of $\beta_t$ parameters, called a \textit{variance schedule}.
The variance schedule $\beta_t$ is treated as a set of hyperparameters of the DDPM model.
\begin{equation}
    q(\vx_t | \vx_{t-1}) := \mathcal{N}\left( \vx_t ; \sqrt{1 - \beta_t} \vx_{t-1}, \beta_t \mathbf{I} \right)
        \label{eq:ddpm_fwd_step}
\end{equation}

A notable property of the Gaussian diffusion process is the existence of the closed-form expression for sampling $\vx_t$ directly
from $\vx_0$ at an arbitrary time $t$, bypassing the whole iterative chain.
Using the notation $\alpha_t := (1 - \beta_t)$ and $\bar \alpha_t := \prod_{s=1}^{s=t} \alpha_s$, it shows that $\vx_t$ can be sampled from $\vx_0$ according to \autoref{eq:ddpm_fwd_jump}.
\begin{equation}
    \vx_t = \sqrt{\bar\alpha_t} \vx_0 + \sqrt{1 - \bar\alpha_t} \veps
        \label{eq:ddpm_fwd_jump}
\end{equation}
where $\veps \sim \stdnorm$.

DDPM approximates the reverse process through a neural network $\veps_\theta(\vx_t, t)$~\cite{ho2020denoising}.
This neural network is trained to predict the noise $\veps$, added in the forward process~(\autoref{eq:ddpm_fwd_jump}),
with a simple $L_2$ loss function~(\autoref{eq:ddpm_loss})
\begin{equation}
    \mathcal{L}(\theta) := \mathbb{E}_{\vx_0, t, \epsilon}
        \left[ \left\|
              \veps
            - \veps_\theta\left(
                \sqrt{\bar\alpha_t} \vx_0 + \sqrt{1 - \bar\alpha_t} \veps, t
              \right)
        \right\|_2^2 \right]
        \label{eq:ddpm_loss}
\end{equation}

Given the neural network $\veps_\theta(\vx_t, t)$, it can be demonstrated that the reverse step is described by~\autoref{eq:ddpm_reverse_step}.
\begin{equation}
    \vx_{t-1} = \frac{1}{\sqrt{\alpha_t}} \left(
        \vx_t - \frac{1 - \alpha_t}{\sqrt{1 - \bar\alpha_t}} \veps_\theta(\vx_t, t)
    \right)
    + \sigma_t \mathbf{z}
    \label{eq:ddpm_reverse_step}
\end{equation}
where $\mathbf{z} \sim \stdnorm$ and $\sigma_t^2 = \tilde \beta_t := (1 - \bar \alpha_{t-1})/(1 - \bar \alpha_t) \beta_t$.
Experimentally, however, DDPM has shown~\cite{ho2020denoising} little difference between using $\sigma_t^2 = \tilde \beta_t$ and $\sigma_t^2 = \beta_t$.
Therefore, we set $\sigma_t^2 = \beta_t$ in this work.

The full DDPM model training (forward process) and data generation (reverse process) procedures are summarized in \autoref{tab:ddpm_train_sample}.
\begin{table*}
\begin{minipage}[t]{0.46\textwidth}
\begin{algorithm}[H]
    \centering
    \caption{DDPM Training}
    \label{alg:ddpm_train}
    \begin{algorithmic}[1]
        \Repeat
            \State $\vx_0 \sim p_\text{data}$
            \State $t \sim \text{Uniform}(\{1, ..., T\})$
            \State $\veps \sim \stdnorm$
            \State Perform Gradient Descent Step on
                \par\qquad
$\nabla_\theta \|
      \veps
    - \veps_\theta \left(
          \sqrt{\bar\alpha_t} \vx_0
        + \sqrt{1 - \bar\alpha_t} \veps,
        t
    \right)
\|^2$            
        \Until {converged}
    \end{algorithmic}
\end{algorithm}
\end{minipage}
\hspace{.8cm}
\begin{minipage}[t]{0.46\textwidth}
\begin{algorithm}[H]
    \centering
    \caption{DDPM Sampling}
    \label{alg:ddpm_sampling}
    \begin{algorithmic}[1]
        \State $\vx_T \sim \stdnorm$
        \For{$t = T, \; ... \;, 1$}
            \State $\vz \sim \stdnorm \; \text{if} \; t > 1, \; \text{else} \; \vz = 0$
            \State
$
    \vx_{t-1} = \frac{1}{\sqrt{\alpha_t}} \left(
            \vx_t - \frac{1 - \alpha_t}{\sqrt{1 - \bar\alpha_t}} \veps_\theta(\vx_t, t)
        \right) + \sigma_t \vz
$
        \EndFor \\
        \Return{$\vx_0$}
    \end{algorithmic}
    \vspace{0.17cm}
\end{algorithm}
\end{minipage}
    \caption{DDPM training algorithm (left) and DDPM sampling algorithm (right) proposed in Ref.~\cite{ho2020denoising}
    }
    \label{tab:ddpm_train_sample}
\end{table*}

\subsection{Factors Influencing DDPM Model Performance}

There are several hyperparameters that influence the DDPM model performance.
The first is the number of diffusion steps $T$.
Empirically, larger $T$ is usually associated with a higher quality of the generated data~\cite{nichol2021improved,song2020denoising}.
The second is the variance schedule $\beta_t$.
The reference DDPM implementation~\cite{ho2020denoising} uses a linear schedule ($\beta_t \propto t$).
However, the improved DDPM models (iDDPM)~\cite{nichol2021improved} have discovered that nonlinear variance schedules may perform better.
Another important factor is the neural network architecture $\veps_\theta(\vx_t, t)$.
iDDPM has demonstrated that a few modifications to the DDPM network architecture may improve the DDPM model performance.

DDPM models are capable of generating very high-quality data, yet their exceptional performance comes at a high computational cost during model inference. 
According to the sampling algorithm (\autoref{alg:ddpm_sampling}), DDPM requires $T$ sequential neural network evaluations (also known as ``sampling steps'') to generate a single image.
This can make DDPM data generation prohibitively expensive.
To speed up the inference, several approximate sampling algorithms have been developed for the DDPM model.
Two of the simplest ones are the iDDPM~\cite{nichol2021improved} and DDIM~\cite{song2020denoising} sub-sampling algorithms.
Both algorithms allow one to perform a DDPM network evaluation at a smaller number of intermediate time steps. 
The smaller number of sampling steps enables a faster model inference at the cost of a small performance degradation.
By default, we do not use the fast sub-sampling algorithms in this work, but we investigate the trade-offs between speed and performance in Section~\ref{sec:discussion}.

\subsection{DDPM Model Training}

This effort relies on a reference implementation~\cite{iddpm_repo} of the diffusion network $\veps_\theta(\vx_t, t)$ from the iDDPM work~\cite{nichol2021improved}. To find the optimal network configuration, we perform a sweep over several network parameters, including varying the depth/width of the model and using attention layers and scale-shift modulation. 

Next, we experiment with using linear and cosine variance schedulers. The cosine variance schedulers result in poor-quality generated samples. Therefore, we use the linear variance scheduler throughout this work.
For the linear scheduler, we conduct a sweep over the $\beta_T$ parameter to determine the best-performing values.

To further improve the model's quality, we vary the number of diffusion steps $T$ and training epochs. The DDPM model performance responds positively to increasing both parameters. We set the number of diffusion steps to $T = \bignum{8000}$, sampling steps to \bignum{8000}, and training epochs to 500 as default because these values provide good generation quality while being small enough for efficient inference. The DDPM model response to the number of training epochs is described in Section~\ref{sec:results}.

Additionally, we have determined the best generation performance is achieved when the training data are normalized to the logarithmic scale.
To perform such a normalization, we first clip the minimal data values by $\SI{1e-3}{GeV}$ (minimum detector resolution) then apply natural logarithm normalization.

Finally, to assess the training stability, each DDPM model configuration is retrained with five random seeds. The variance of the model performance with different seeds is used to judge the training stability (cf.~Section~\ref{sec:results}). 
Additional details, including the final network configuration and training procedure, can be found in Appendix~\ref{sec:app_ddpm_train_details}.

\begin{figure*}[t!]
    \centering
    \begin{tikzpicture}
        \node[inner sep=0] (cent0) at (0, 0) {\includegraphics[width=0.98\textwidth]{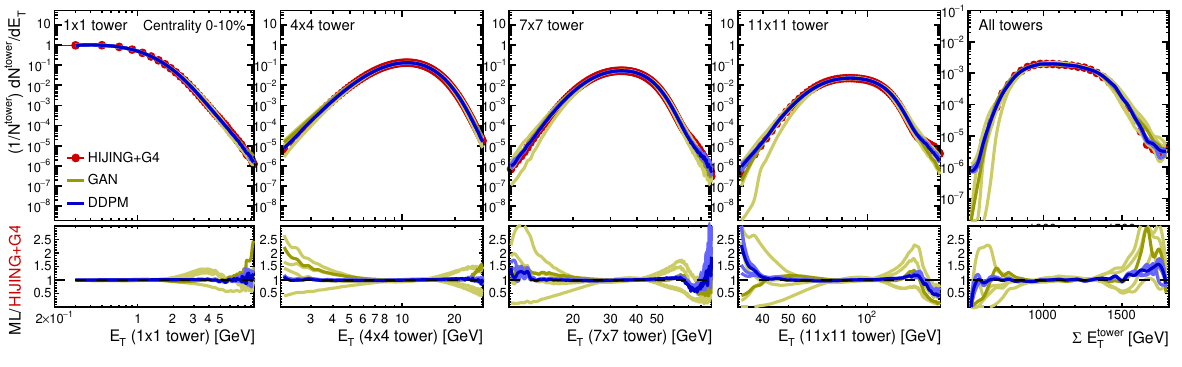}};
        \node[inner sep=0, anchor=south west] (cent0_title) at (cent0.north west) {A.~Calorimeter tower \et distributions for $0$--$10\%$ centrality};
        \node[inner sep=0, anchor=north] (cent4) at ($(cent0.north)!1.1!(cent0.south)$) {\includegraphics[width=0.98\textwidth]{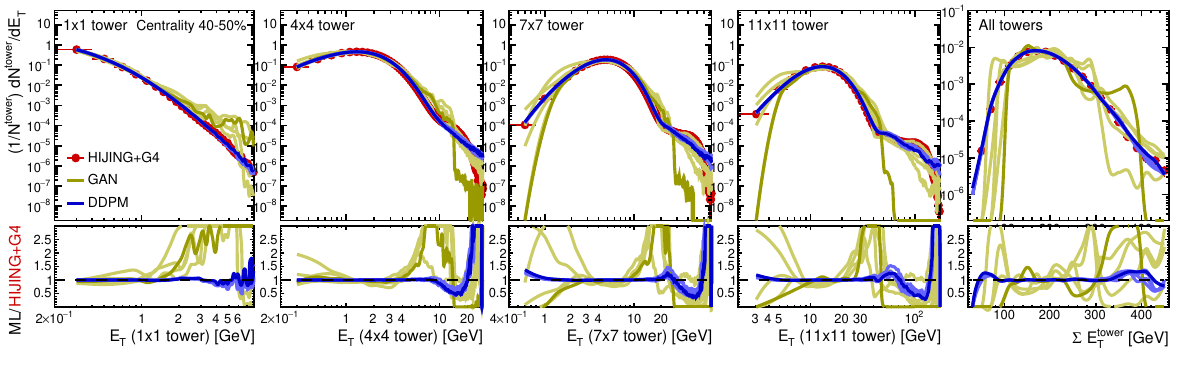}};
        \node[inner sep=0, anchor=south west] (cent4_title) at (cent4.north west) {B.~Calorimeter tower \et distributions for $40$--$50\%$ centrality};
    \end{tikzpicture}
    \caption{Calorimeter tower \et distributions for \hijinggeant (red circle), DDPM (blue lines), and GAN (yellow lines) in the $0$--$10\%$ centrality range (A) and the $40$--$50\%$ centrality range (B). The first four columns represent \et of different tower areas: $1\times1$, $4\times4$, $7\times7$, and $11\times11$. The rightmost column is the distribution of total \et per event (\sumettower). Bottom panels show the ratio of ML models to \hijinggeant. The five different lines for GAN and DDPM represent varied random seeds. 
    }
    \label{fig:etDist}
\end{figure*}
\subsection{GAN Baselines}

GANs~\cite{goodfellow2020generative} present an alternative family of generative models.
Similar to the diffusion models, they offer the ability to synthesize high-quality data but with much faster inference times. In this work, we also explore GANs' ability to generate samples for comparison with the DDPM model.

For the GAN baselines, we choose a widely used DCGAN model~\cite{radford2015unsupervised} and construct its generator and discriminator networks out of five DCGAN stages (with $2^{6}$, $2^{7}$, $2^{8}$, $2^{9}$, and $2^{10}$ channels per stage, respectively). 
The output of a DCGAN generator is an image of size $(64, 64)$ pixels. Because the shape of the generated image does not match that of the towers $(24, 64)$, we crop the image to the required shape.
Similarly, the DCGAN discriminator expects an input image of shape $(64, 64)$.
To make the discriminator work on towers of $(24, 64)$, we embed them into a larger $(64, 64)$ image, initially filled with zeros. 

Similar to the DDPM models, we find that a logarithmic data normalization gives better performance compared to an unaltered case.
Likewise, all the final DCGAN configurations are retrained with five random seeds to assess their stability. The full training details of the GAN baselines are outlined in Appendix~\ref{sec:app_gan_train_details}. 
Despite numerous tuning attempts, the quality of the GANs-generated samples in this work is too poor for it to be considered a viable generation methodology. 
GANs also exhibit a high degree of training instability further limiting their applicability for high-fidelity generation.
Additional details regarding the performance of the GANs and their comparison to the DDPM are discussed in Section~\ref{sec:results}.

\begin{figure*}[t]
    \centering
    \begin{tikzpicture}
        \node[inner sep=0] (cent0) at (0, 0) {\includegraphics[width=0.98\textwidth]{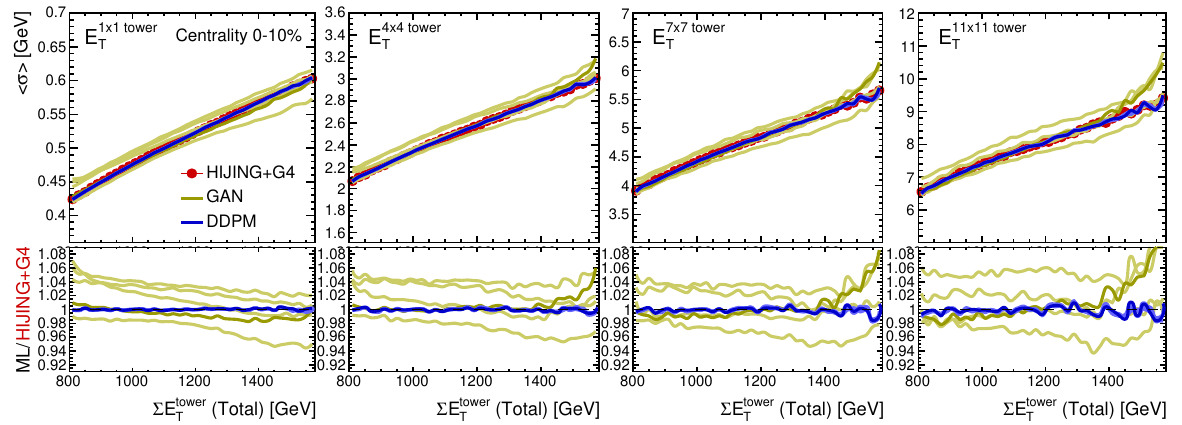}};
        \node[inner sep=0, anchor=south west] (cent0_title) at (cent0.north west) {A.~Average standard deviation of tower \et as a function of \sumettower\ for the $0$--$10\%$ centrality};
        \node[inner sep=0, anchor=north] (cent4) at ($(cent0.north)!1.1!(cent0.south)$) {\includegraphics[width=0.98\textwidth]{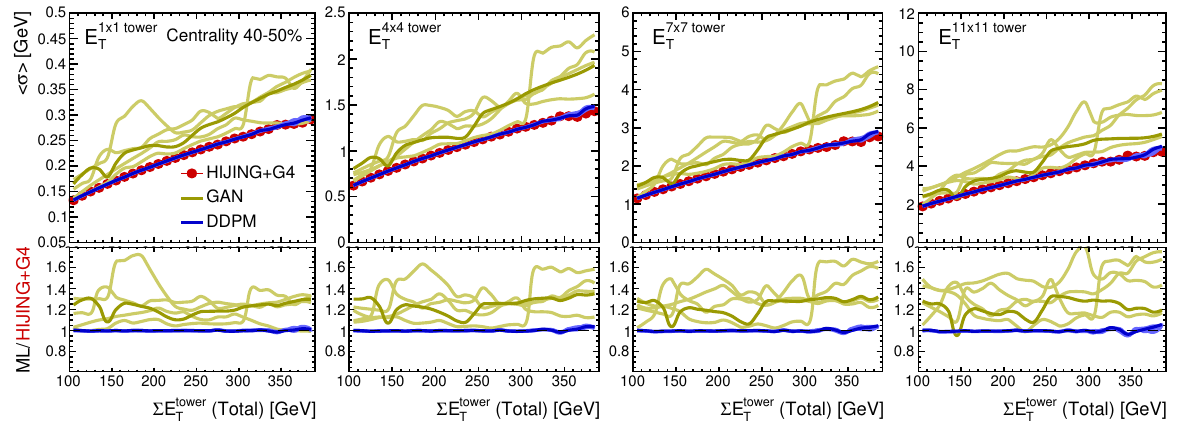}};
        \node[inner sep=0, anchor=south west] (cent4_title) at (cent4.north west) {B.~Average standard deviation of tower \et as a function of \sumettower\ for the $40$--$50\%$ centrality};
    \end{tikzpicture}
    \caption{Average standard deviation (\avgStdEt) of calorimeter tower \et as a function of \sumettower\ for \hijinggeant (red circle), DDPM (blue lines), and GAN (yellow lines) in the $0$--$10\%$ centrality range (A) and the $40$--$50\%$ centrality range (B). Bottom panels show the ratio of ML models to \hijinggeant.
     The five different lines for GAN and DDPM represent varied random seeds. 
    }
    \label{fig:avgStdEt_vs_towerEt}
\end{figure*}

\section{Results}
\label{sec:results}

The distribution of deposited calorimeter tower energies and its event-by-event fluctuation are compared between the \hijinggeant and generated events by DDPMs and GANs to evaluate the efficacy of the ML models. A total of \bignum{100000} events are utilized for each dataset in this evaluation. The \hijinggeant events used for evaluation is independent from \hijinggeant events used for training.
\autoref{fig:etDist}A shows transverse energy (\et) distributions deposited in various calorimeter tower areas ($1\times1$, $4\times4$, $7\times7$, and $11\times11$) and the distribution of the total \et of all towers (\sumettower) in the $0\,$--$10\%$ centrality range. The tower area is defined as the region occupied by the array of towers in the $\eta$ and $\phi$ direction. 
For example, a $7\times7$ tower area consists of 49 towers. 
Both the DDPM and GAN models are trained five times with different random seeds to test stability. The DDPM has robust consistency across multiple seeds, whereas the GAN model shows significant fluctuations when varying seeds. The DDPM also offers a significantly more accurate description of \hijinggeant, typically within a few percentage points of accuracy where the distribution is populated, while the GAN fails to replicate the \hijinggeant, which is particularly evident at the distribution tails.

\autoref{fig:etDist}B shows the calorimeter tower \et distributions in the $40$--$50\%$ centrality range. 
Similar to the $0$--$10\%$ case, the DDPM outperforms the GAN in both the stability and better description of \hijinggeant. However, the DDPM starts to deviate from the \hijinggeant at higher \et regions, where the distribution displays a non-Gaussian structure. This structure likely arises from pileup events and minijets generated in HIJING, which occur rarely. The probability of encountering this high \et tail at around $4\times4$ tower \et of 15 GeV is approximately $10^{-5}$, with only around $100$ occurrences contributing to this distribution in this evaluation. 

In addition to the overall energy deposition, we examine energy fluctuations. The standard deviation of transverse energies (\stdEt) is calculated per event for different tower size configurations ($1\times1$, $4\times4$, $7\times7$, and $11\times11$ towers). The average \stdEt, \avgStdEt, is shown as a function of \sumettower\ in~\autoref{fig:avgStdEt_vs_towerEt}A and~\autoref{fig:avgStdEt_vs_towerEt}B for the $0$--$10\%$ and $40$--$50\%$ centrality ranges, respectively. Similar to the overall energy distributions, the DDPM provides significantly improved performance than the GAN and closely aligns with the HIJING across various tower configurations. This highlights the robust descriptive capability of the DDPM model.

\section{Discussion}
\label{sec:discussion}

\begin{figure*}[hbtp!]
    \centering
    \begin{tikzpicture}
        \node[inner sep=0] (epoch) at (0, 0) {\includegraphics[width=0.98\textwidth]{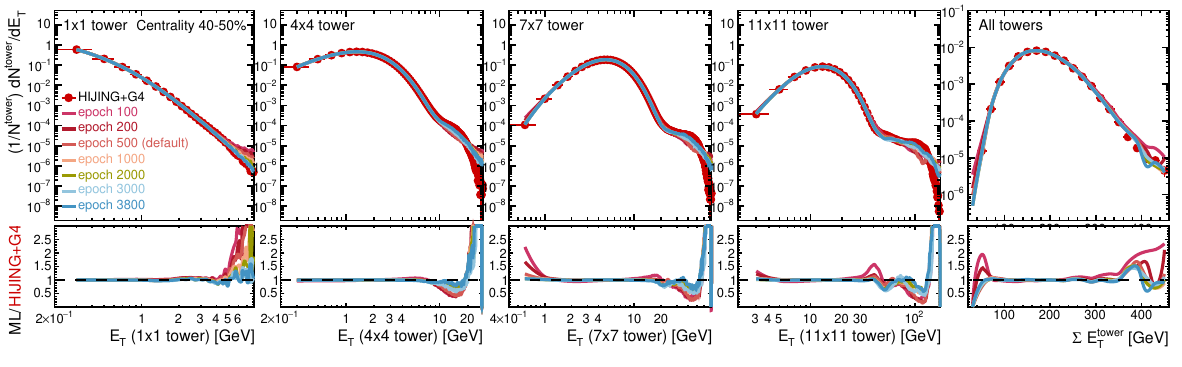}};
        \node[inner sep=0, anchor=south west] (epoch_title) at (epoch.north west) {A.~Calorimeter tower \et distributions with a different number of training epochs of the DDPM};
        \node[inner sep=0, anchor=north] (step) at ($(epoch.north)!1.1!(epoch.south)$) {\includegraphics[width=0.98\textwidth]{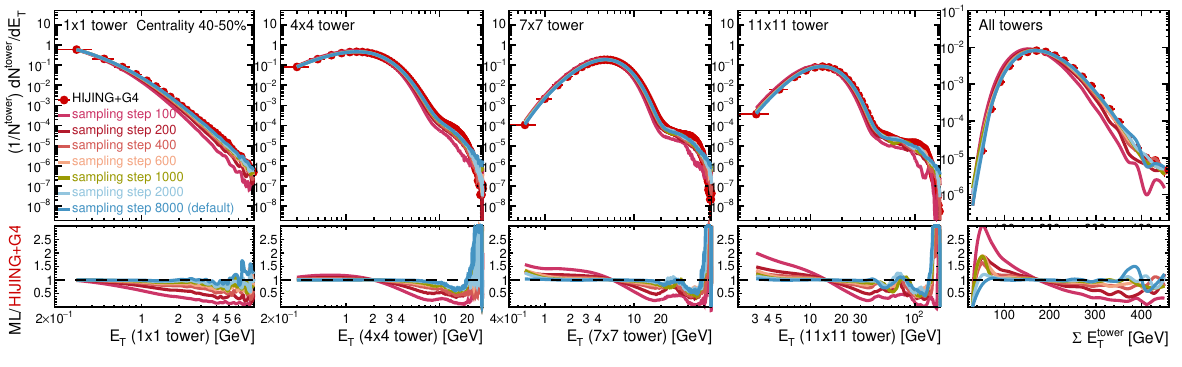}};
        \node[inner sep=0, anchor=south west] (step_title) at (step.north west) {B.~Calorimeter tower distributions with a different number of sampling steps of the DDPM};
    \end{tikzpicture}
    \caption{Calorimeter tower \et distributions in the $40$--$50\%$ centrality range with different numbers of training epochs (A) and sampling steps (B) of the DDPM. All curves in A are produced with 8000 sampling steps. All curves in B are produced with 4000 epochs.}
    \label{fig:epoch_step}
\end{figure*}

Machine learning algorithms tend to work suboptimally when the training data are scarce. This is a significant problem in the context of high-energy heavy-ion collisions as the instances of large energy depositions are rare but essential for the overall fidelity of the simulation. The rare and large energy depositions often include high-energy phenomena, such as jets and photons, which are of scientific interest. 
To alleviate the problem and enhance fidelity, we increase the number of training epochs for the network to learn the distribution more accurately in less dense regions. \autoref{fig:epoch_step}A shows that by increasing the training epochs, the DDPM works progressively better at the higher energy region of the distribution. 
It is important to note that increasing the number of epochs also prolongs the training time. In this work, each training session took 4 hours and 44 minutes per 100 epochs with the duration increasing linearly as the number of epochs rises.
Therefore, there is a trade-off between the training time and quality with which rare features are reproduced.

The number of DDPM sampling steps also affects the time required to generate events. To explore the trade-offs between fidelity and generation speed, we have experimented with a fast sampling strategy proposed by iDDPM~\cite{nichol2021improved}.
\autoref{fig:epoch_step}B shows the impact of reducing the number of sampling steps from the default value of 8000 to 100.
The DDPM model shows a steep improvement in generation quality as the number of sampling steps increases from $100$ to \bignum{1000}.
Additional increases in the number of sampling steps bring progressively diminishing returns with \bignum{2000} sampling steps providing similar performance to \bignum{8000}.
This observation indicates that the generation speed can be improved by a factor of $4$ to $8$ without significant loss of quality. 
To further explore the possibilities of faster generation, we attempt to reevaluate the trained DDPM models in the DDIM mode~\cite{song2020denoising}. However, DDIMs do not provide sufficient generation quality to be a viable alternative to the DDPM model. Additional details about generation with the DDIM model are discussed in Appendix~\ref{sec:app_ddim}.

The average time to produce one event with HIJING and simulate it with Geant4 takes approximately 40 minutes using a single CPU core. Meanwhile, utilizing an NVIDIA RTX A6000 GPU, the GAN and DDPM with the denoising step of $8000$ require only about $0.42$ milliseconds and $1.34$ seconds, respectively. This represents a speedup of \bignum{5700000} and \bignum{1800} times faster than conventional event simulation. For a fair comparison between CPU and GPU, considering a $32$-core CPU equivalent to a GPU, the GAN and DDPM still offer a speedup on the order of \bignum{100000} and $100$, respectively. While the GAN demonstrates greater speed, the DDPM maintains high fidelity in describing the ground truth, which is crucial for scientific purposes.

It is important to note that traditional simulation tools, such as event generators and Geant4, have a wide range of physics-driven model parameters that allow for the study of systematic uncertainties due to model parameter dependencies. The algorithm presented in this work does not support such a feature. Instead, it is designed to faithfully replicate a reference dataset in order to generate a much larger event sample efficiently. While it is possible to incorporate model parameters into models such as DDPM, doing so would require a much larger training set covering the parameter space, which is beyond the scope of this work.

In applications, there is a need for a large number of full-detector simulation events, such as those for heavy-ion collisions at RHIC and beam background events for the EIC detector. Rare signal events can then be embedded in these simulations. We envision using a relatively modest number (at the level of millions) of events simulated through Geant4 to train the diffusion model. Once trained, the model can be used to accelerate the production of much larger samples (at the level of billions) for signal embedding. 

While the generation of samples using DDPM is much more time-efficient compared to traditional methods, caution should be exercised regarding their limitations in regions with sparse data. As discussed, fidelity can be improved by increasing epochs and sampling steps, optimizing them to address limitations in data description performance. This limitation may be less significant in certain cases, such as when embedding jet signal samples into a heavy ion event sample, where the sparse data regions are supplemented by the abundant jet signal samples.

In addition to the tower \et distributions and their fluctuations, there are more ways to characterize heavy-ion collisions and evaluate the performance of the generation, including tower correlations, azimuthal anisotropy and resonance (e.g. $\pi^{0}$) reconstruction, which is a subject for future work with larger statistics.

\section{Summary}

Simulating particle interactions with detectors in nuclear experiments is highly computationally demanding.  
This study introduces the first application of generative AI as an alternative to traditional simulation methods for full-event, whole-detector simulation of heavy-ion collisions. We employ both GANs and DDPM to generate events and compare them with the traditional method of HIJING+Geant4 simulation in detail. We find that the DDPM shows promise for high-fidelity and stable simulation in this application while maintaining orders of magnitude speed gain when compared with the traditional methods. This work demonstrates diffusion models are potentially useful tools for future full-detector, whole-event simulation campaigns in the nuclear physics field, such as experiments at the Relativistic Heavy Ion Collider, Large Hadron Collider, and future EIC.

Source code is available in~\cite{ddpm_repo_ours}.

\section*{Acknowledgment}

We thank the sPHENIX collaboration for access to the simulated dataset, which was used in the training and validation of our algorithm. We also thank for valuable interaction with the sPHENIX collaboration on this work.

This manuscript has been authored by employees of
Brookhaven Science Associates, LLC under Contract No.
DE-SC0012704 with the U.S. Department of Energy. The
publisher by accepting the manuscript for publication acknowledges that the United States Government retains a
non-exclusive, paid-up, irrevocable, world-wide license to
publish or reproduce the published form of this manuscript, or
allow others to do so, for United States Government purposes.

\appendix

\section{Training Details}

This appendix summarizes various training details for the DDPM and GAN models used in the main paper.

\subsection{DDPM Training Details}
\label{sec:app_ddpm_train_details}
\begin{figure*}[t!]
    \centering
    \includegraphics[width=0.95\textwidth]{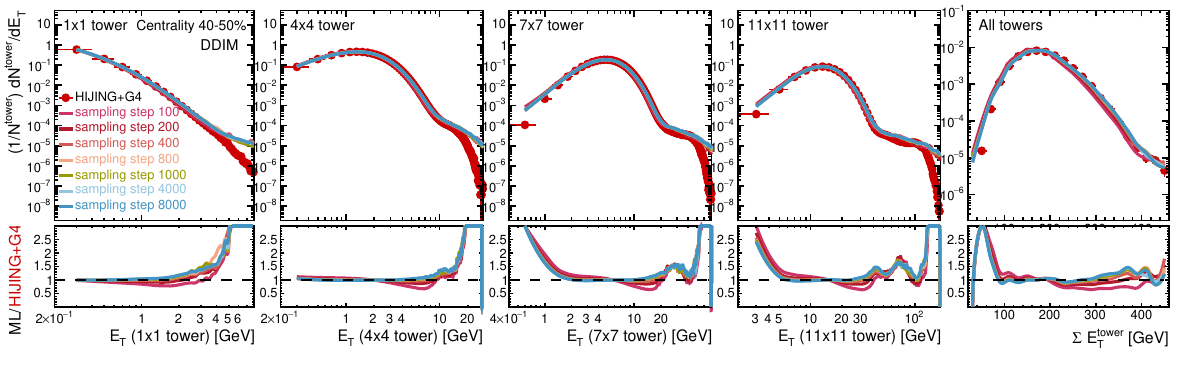}
    \caption{
    Calorimeter tower \et distributions in the 40--50\% centrality range with a different number of DDIM mode sampling steps. 
    }
    \label{fig:DDIM}
\end{figure*}
This work relies on a reference iDDPM implementation~\cite{iddpm_repo} of the diffusion network but changes its configuration.
Our neural architecture is built from three U-Net encoder-decoder stages~\cite{ronneberger2015u} (with 32, 64, and 128 channels per stage, respectively), each of which comprises two ResNet blocks~\cite{he2016deep}.
This architecture is chosen as it gives good generation performance while not being overly complex.
The iDDPM model allows one to incorporate attention layers into the U-Res-Net architecture.
However, we did not observe any sizeable performance improvements with the inclusion of attention layers.
Therefore, our network architecture is fully convolutional and does not use any attention layers.
The scale-shift time modulation produces a slightly different effect, depending on the dataset. The DDPM shows better performance for the centrality 40--50\% range with the scale-shift modulation, but the opposite is true for the centrality 0--10\% dataset.

To further investigate performance of the DDPM models, we performed a hyperparameter sweep over the end value of the linear variance schedule $\beta_T \in (0.01, 0.20)$. Our findings indicate that the DDPM performs best with $\beta_T = 0.02$ for the centrality 40--50\% range and $\beta_T = 0.10$ for the centrality 0--10\% range.

As the best DDPM model configurations differ between the datasets, we used different setups per dataset in the main paper:
\begin{enumerate}[noitemsep, topsep=8pt, label=(\roman*)]
    \item Centrality 0--10\%:  No scale-shift modulation, $\beta_T = 0.02$
    \item Centrality 40--50\%: Scale-shift modulation, $\beta_T = 0.10$
\end{enumerate}

Overall, the DDPM models are trained with a batch size of $128$. For consistency between different datasets, we limit the number of training steps per epoch to \bignum{2000}.
For the default training (Figures~\ref{fig:etDist},~\ref{fig:avgStdEt_vs_towerEt}), the DDPM models are trained for $500$ epochs.
To perform the additional investigations (Figures~\ref{fig:epoch_step},~\ref{fig:DDIM}) on the centrality 40--50\% sample, we train the DDPM models for \bignum{4000} epochs (unless otherwise indicated).
The training was performed with the Adam optimizer~\cite{kingma2014adam} having a constant learning rate of $1 \times 10^{-4}$.

\subsection{GAN Training Details}
\label{sec:app_gan_train_details}
The GANs used in this work are based on DCGAN architecture~\cite{radford2015unsupervised} and try to follow its training procedure closely. However, we found the default DCGAN training setup results yields significantly poor performance.
To improve the situation, we experimented with using alternative loss functions and a gradient penalty (GP) term~\cite{gulrajani2017improved}.

Specifically, we conducted hyperparameter sweeps over the choice of the loss function and use of the GP term~\cite{gulrajani2017improved}.
For the loss function, we explored using the traditional binary cross entropy (BCE) loss function~\cite{goodfellow2020generative} (default DCGAN setup), WGAN-GP loss~\cite{gulrajani2017improved}, and least squares GAN (LSGAN) loss~\cite{mao2017least}.
For the GP configuration, we considered not using any gradient penalty (default DCGAN setup), using the original GP~\cite{gulrajani2017improved}, and using the improved zero-centered GP term~\cite{thanh2019improving}.
For each of the GP configurations, we also experimented with its magnitude $\lambda_\text{GP}$ from $\{ 1, 10 \}$.

Our findings indicate that the following configurations perform the best per dataset:
\begin{enumerate}[noitemsep, topsep=8pt, label=(\roman*)]
    \item Centrality 0--10\%: Loss: WGAN-GP with the original GP of magnitude $\lambda_\text{GP} = 1$.
    \item Centrality 40--50\%: Loss: LSGAN with the original GP of magnitude $\lambda_\text{GP} = 10$
\end{enumerate}

These configurations are used in the main paper. All DCGAN models are trained with a batch size of $128$ (following the original DCGAN setup) for \bignum{1000000} steps.
We found no benefit in continuing the training further. 
For optimization, we used the Adam optimizer~\cite{kingma2014adam} with the learning rate of $1 \times 10^{-4}$.

\section{DDPM Evaluation in DDIM Mode}
\label{sec:app_ddim}

The DDIM~\cite{song2020denoising} model provides an alternative formulation of the diffusion process. Unlike the DDPM model, the reverse process of DDIM is fully deterministic.
This allows for establishing a 1-to-1 correspondence between the noise $\stdnorm$ and the data $p_\text{data}$ distributions, which offers many interesting applications related to the modifications in the noise space~\cite{song2020denoising,preechakul2022diffusion}.
The DDIM also offers an alternative way of approximate inference that may outperform DDPM sub-sampling when the number of sampling steps is small~\cite{song2020denoising}.

The DDIM model's training objective is constructed to match that of the DDPM model.
This property permits us to reuse a trained DDPM model and evaluate it in a DDIM mode.
\autoref{fig:DDIM} shows the results of evaluating the DDPM models from the main paper in the DDIM mode.

Comparing DDIM inference (\autoref{fig:DDIM}) with the corresponding DDPM inference (\autoref{fig:epoch_step}) of the same DDPM model, we observe a rather significant difference in the quality of the generated data.
In particular, the DDIM performs poorly in the tails of the tower \et distributions, irrespective of the number of sampling steps.

The especially poor performance of the DDIM model is relatively surprising given that it works well on natural images~\cite{song2020denoising}.
It is possible that the relatively better DDPM performance in the low-statistics region is due to the stochastic nature of the DDPM denoising process.
The stochasticity of the denoising process may help to perturb initial samples enough to generate low-frequency data correctly.
It is also possible that the DDIM's underperformance stems from our models being optimized for DDPM performance, which may not align with optimizing for DDIM performance. 
Additional studies are necessary to differentiate the effects of DDIM determinism from the lack of optimization.

\bibliography{main}

\end{document}

%% file: macro.tex
\newcommand{\sphenix}{sPHENIX\xspace}
\newcommand{\snn}{\ensuremath{\sqrt{\mathrm{s}_{\mathrm{NN}}}\xspace}}
\newcommand{\deta}{\ensuremath{\Delta\eta}}
\newcommand{\dphi}{\ensuremath{\Delta\phi}}
\newcommand{\et}{\ensuremath{E_\mathrm{T}}\xspace}
\newcommand{\sumettower}{\ensuremath{\Sigma E_\mathrm{T}^\mathrm{tower}}}
\newcommand{\stdEt}{\ensuremath{\sigma^{E_\mathrm{T}}}\xspace}
\newcommand{\avgStdEt}{\ensuremath{\langle \sigma^{E_\mathrm{T}} \rangle}\xspace}
\newcommand{\hijinggeant}{HIJING+G4\xspace}

\newcommand{\stdnorm}{\mathcal{N}(\mathbf{0}, \mathbf{I})\xspace}
\newcommand{\vx}{\mathbf{x}\xspace}
\newcommand{\vz}{\mathbf{z}\xspace}
\newcommand{\veps}{\mathbf{\epsilon}\xspace}
\newcommand{\bignum}[1]{\num[group-separator={,}]{#1}} 

\newcommand{\grass}{sample\xspace}

\newcommand{\algorithmautorefname}{Algorithm} 